\documentclass[aps,preprint,nofootinbib,floatfix,superscriptaddress]{revtex4-1}

\usepackage{amsmath,amssymb,amsfonts,color,graphicx,graphics,latexsym,placeins,epsfig,multirow,bm,physics,bbold,enumitem,array}

\usepackage{array}
\newcolumntype{P}[1]{>{\centering\arraybackslash}p{#1}}
\usepackage{footmisc}
\usepackage[compatibility=false]{caption}
\usepackage{subcaption}
\usepackage[font=small,labelfont=bf]{caption}
\usepackage{comment}
\usepackage{bbold}
\usepackage{enumitem}
\usepackage{braket}
\setdescription{leftmargin=8pt}
\usepackage{hyperref}
\hypersetup{
	colorlinks=true,       % false: boxed links; true: colored links
	linkcolor=black,          % color of internal links
	citecolor=black,        % color of links to bibliography
	urlcolor=black           % color of external links
}
\usepackage{natbib}
\setcitestyle{square, comma, numbers, sort&compress}
\usepackage{amsmath,amssymb}

\usepackage{booktabs} 
\usepackage[compatibility=false]{caption}
\usepackage[font=small,labelfont=bf]{caption}
\usepackage{varioref}
\usepackage[active]{srcltx}
\usepackage{fancybox}
\usepackage[dvipsnames]{xcolor}
\usepackage{soul}

\newcommand{\ba}{\begin{align}}
\newcommand{\ea}{\end{align}}

\def\3nab{\tilde{\nabla}}

\def\be {\begin{equation}}
\def\ee {\end{equation}}
\def\ba {\begin{eqnarray}}
\def\ea {\end{eqnarray}}

\newcommand{\sfr}[2]
{{\textstyle\frac{#1}{#2}}}

\newcommand{\barray}{\begin{array}}
\newcommand{\earray}{\end{array}}

\newcommand{\bea}{\begin{eqnarray}}
\newcommand{\eea}{\end{eqnarray}}

\begin{document}
%%%%%%%%%%%%%%% AUTHOR'S NAMES AND AFFILIATIONS %%%%%%%%%%%%%%%%%%
%
%\begin{comment}
\title{\Large \bf Finite 4-D Gauss-Bonnet quantum corrections from matter-graviton interactions in a curved background}

\author{Apurv Keer}
\email{apurv.keer@iitb.ac.in}

\author{S. Shankaranarayanan}
\email{shanki@iitb.ac.in}

\affiliation{Department of Physics,  Indian Institute of Technology Bombay, Mumbai 400076, India}
%%%%%%%%%%%%%%%%%%%%%%%%%%% ABSTRACT  %%%%%%%%%%%%%%%%%%%%%%%%%%

\begin{abstract}
The Glavan-Lin proposal for 4D Einstein-Gauss-Bonnet (EGB) gravity introduces a singular dimensional scaling to bypass Lovelock's theorem, though its fundamental origin remains debated. In this work, we demonstrate that this specific dimension-dependent scaling naturally emerges from the one-loop self-energy corrections of gravitons. By employing real-space techniques to evaluate graviton interactions with minimally coupled scalar and electromagnetic fields \emph{in a de Sitter background}, we show that the $1/(D-4)$ pole from dimensional regularization naturally yields a term proportional to the Gauss-Bonnet invariant. Importantly, because the Gauss-Bonnet density is a total derivative in four dimensions, this combination yields a strictly finite quantum correction analogous to the conformal anomaly, rather than a divergent counterterm. We confirm that the true ultraviolet divergences are strictly renormalized by standard quadratic curvature counterterms, specifically the Weyl-squared and Ricci-scalar-squared invariants. Our results indicate that the Glavan-Lin scaling is \emph{not an ad-hoc} classical limit, but effectively captures a finite, loop-induced quantum feature of the effective action. We discuss the implications of these finite, background-dependent corrections in the early-Universe and their consequences in the strong gravity regime.


\noindent\textbf{Corresponding author:} Apurv Keer

\end{abstract}

\maketitle

\section{Introduction}

General Relativity (GR) stands as one of the most successful theories in modern physics. Evolving as a natural extension of special relativity, it describes gravity not as a force, but as the dynamic geometry of spacetime itself. This geometric perspective --- where the metric and connection are dynamical variables --- has been validated by a century of precision tests, from the deflection of light to the detection of gravitational waves. However, the very success of GR highlights its limitations~\cite{Shankaranarayanan:2022wbx}. The existence of spacetime singularities, such as those at the centers of black holes and the Big Bang, signals a breakdown of the classical description~\cite{Mandal:2025xuc}. Furthermore, on cosmological scales, the observed acceleration of the universe suggests either the presence of dark energy or the need for infrared modifications to gravity. Perhaps most critically, the quantization of GR remains elusive; in $D$ dimensions, the gravitational constant $G_D$ has a mass dimension of $(2-D)$, rendering Einstein-Hilbert gravity non-renormalizable by standard perturbative methods~\cite{Deser:1957zz,Donoghue:1997hx,Shomer:2007vq}.

To address these challenges, modified gravity theories have been extensively explored~\cite{Shankaranarayanan:2022wbx,Mandal:2025xuc}. Lovelock's theorem provides a rigorous classification for such modifications, stating that in 4-D, the Einstein field equations are the unique second-order equations of motion derived from a diffeomorphism-invariant action~\cite{Lovelock:1972vz,Padmanabhan:2013xyr}. Higher-order curvature terms, such as the Gauss-Bonnet (GB) invariant $\mathcal{G} = R^2 - 4R_{\mu\nu}^2 + R_{\mu\nu\rho\sigma}^2$, are purely topological in four dimensions and do not contribute to the dynamics. However, in 2020, Glavan and Lin proposed a novel framework to bypass this restriction~\cite{Glavan:2019inb}. They suggested a singular renormalization of the coupling constant, scaling $\alpha \to \alpha/(D-4)$ in $D$ dimensions. By taking the limit $D \to 4$, they argued that the GB term yields a non-trivial contribution to the equations of motion, effectively bypassing Lovelock's theorem. This \emph{4D Einstein-Gauss-Bonnet} gravity has since sparked intense interest in black hole physics and cosmology~\cite{Fernandes:2022zrq}.

Despite its phenomenological promise, the Glavan-Lin proposal has faced significant criticism regarding its mathematical consistency~\cite{Konoplya:2020juj}. Several authors have pointed out that the limit is not continuous and lacks a canonical definition, potentially leading to ill-defined interacting theories or coupling blow-ups~\cite{Gurses:2020ofy, Ai:2020peo, Hennigar:2020lsl, Mahapatra:2020rds}. Specifically, the regularization depends on how the higher-dimensional space is compactified, and alternative consistent realizations often reduce to scalar-tensor theories of the Horndeski class rather than pure gravity~\cite{Fernandes:2022zrq}. Consequently, there is a pressing need to establish whether the $1/(D-4)$ scaling is merely \emph{an ad-hoc} classical trick or if it has a fundamental origin in the quantum structure of the theory.

In a previous work~\cite{Mandal:2024kic}, we addressed this ambiguity by grounding the scaling process in standard QFT renormalization. We demonstrated that the one-loop self-energy correction of gravitons due to matter loops in flat spacetime naturally generates a term proportional to $\mathcal{G}/(D-4)$. Because the Gauss-Bonnet density is a total derivative in four dimensions, this combination yields a strictly finite quantum correction to the effective action. It does not cancel UV divergences, but acts as a finite quantum modification analogous to the trace anomaly. Crucially, the true UV divergences are instead absorbed by standard quadratic curvature counterterms --- specifically the Weyl-squared and Ricci-scalar-squared invariants --- that naturally exhibit a $1/(D-4)$ pole structure in dimensional regularization. This confirms that the \emph{singular limit} effectively captures a natural consequence of semi-classical loop integration when matter fields are integrated out.

However, a robust validation of this proposal requires extending the analysis beyond flat space. This necessitates a shift from momentum-space techniques to real-space (coordinate-space) methods. In flat spacetime, translation invariance allows for a convenient Fourier decomposition, making momentum-space Feynman integrals the standard tool for computing self-energies. In curved spacetime, however, global translation invariance is broken~\cite{DeWitt:1975ys,Birrell:1982ix}. The notion of a particle and the definition of the vacuum become observer-dependent, and the Fourier modes of the fields are no longer simple plane waves~\cite{Fulling:1989nb,Mukhanov:2007zz}. Consequently, the graviton and matter propagators cannot be simply diagonalized in momentum space~\cite{Parker:2009uva}. To probe the ultraviolet (UV) structure of the theory in a background-independent manner, one must employ real-space techniques, such as the point-splitting method or the Schwinger-DeWitt proper time expansion~\cite{Christensen:1976vb, Bunch:1979uk}. These methods focus on the short-distance singularity structure of the Green's function, $G(x, x') \sim \sigma(x,x')^{1-D/2}$, where $\sigma(x,x')$ is the geodesic distance. Since the UV divergences responsible for the renormalization of the gravitational coupling are local phenomena determined by the coincidence limit $x \to x'$, real-space analysis is indispensable. It allows us to isolate the pole in $(D-4)$ purely from the geometric structure of the propagator, without relying on the ambiguities of defining momentum in curved geometry. Furthermore, real-space methods provide a direct link to the local curvature invariants (like the Gauss-Bonnet term) appearing in the effective action~\cite{Barth:1983hb}.

In this work, we extend our real-space analysis to de Sitter (dS) spacetime. {While it is a fundamental theorem of quantum field theory that the structural coefficients of UV divergent counterterms are strictly local and therefore background-independent, evaluating the self-energy in an explicit curved background is indispensable for two reasons: first, to demonstrate the robustness of our coordinate-space regularization scheme away from flat space; and second, to track the background-dependent finite quantum corrections (proportional to the Hubble parameter $H$) that dictate early-universe phenomenology. By computing the one-loop graviton self-energy in the Poincar\'e patch of dS, we verify that the algebraic $1/(D-4)$ pole structure naturally arises in semi-classical gravity, independent of the background geometry. Specifically, we show that integrating out a minimally coupled scalar field yields a finite Gauss-Bonnet quantum correction alongside a Ricci-scalar-squared counterterm.} For the electromagnetic sector, we demonstrate that the one-loop photon self-energy induces the finite Gauss-Bonnet modification alongside a Weyl-squared counterterm, a structure directly linked to the conformal anomaly of the electromagnetic field in curved spacetime.

The remainder of this paper is organized as follows. In Sec.~\ref{sec:background}, we establish the general field-theoretic framework for evaluating the one-loop graviton self-energy induced by matter fields. In Sec.~\ref{sec:oneloop-Minkowski}, we briefly review the momentum-space derivation in flat spacetime before developing the covariant real-space (coordinate-space) formalism. This section demonstrates how the $1/(D-4)$ pole and the associated counterterms can be isolated purely from the short-distance singularity structure of the propagator. In Sec.~\ref{sec:oneloop-scalar-deSitter}, we extend this real-space methodology to a curved de Sitter background, explicitly computing the self-energy corrections for a minimally coupled scalar field and confirming the background-independent emergence of the finite 4D Gauss-Bonnet quantum correction. Sec.~\ref{sec:oneloop-photon-deSitter} generalizes this analysis to the electromagnetic field, detailing the exact gauge-fixing procedure required in de Sitter spacetime and deriving the corresponding Weyl-squared counterterms. In Sec.~\ref{sec:Early-Universe}, we discuss the physical implications of these loop-generated higher-derivative terms and explore their role in driving early-universe acceleration via Starobinsky-type dynamics. We conclude in Sec.~\ref{sec:discussion} by discussing possible implications in the strong gravity regime and proposing a framework to constrain these loop-corrected scattering amplitudes using quantum optical phenomena. In this work, we use natural units $\hbar = c = 1$, and set $\kappa^2 = 8\pi G_D$ with $G_D$ being the $D$-dimensional Newton's constant.

\section{One loop graviton self energy}
\label{sec:background}

We start by expressing the full metric as a perturbation around a fixed background:
\begin{equation}\label{metric_perturbation}
g_{\mu\nu} = \Bar{g}_{\mu\nu} + \kappa h_{\mu\nu}
\end{equation}
where $\Bar{g}_{\mu\nu}$ is the background metric, $h_{\mu\nu}$ is the graviton field with indices raised and lowered by the background metric.  We derive the interaction vertices between the scalar field and dynamical gravitons by expanding the massless, minimally coupled scalar Lagrangian. The total Lagrangian describing pure gravity and the interaction is:
\begin{equation}
\mathcal{L} = \mathcal{L}_{EH} -\frac{1}{2} \sqrt{-g} g^{\mu \nu} \partial_\mu \phi \partial_\nu \phi
\end{equation}
\begin{figure}[h]
    \centering
    \includegraphics[width=0.5\linewidth]{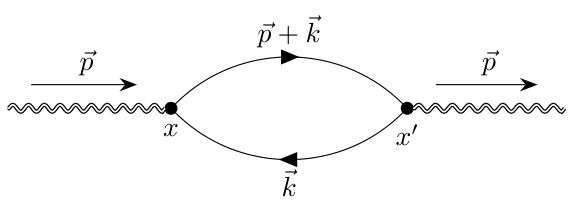}
    \caption{One-loop Feynman diagram describing the self-energy correction to the graviton propagator due to a massless scalar field loop.}
    \label{fig:scalar corr feynman}
\end{figure}
The one-loop scalar contribution to the graviton self-energy, corresponding to the diagram in Fig. \ref{fig:scalar corr feynman}, is given by the expectation value of the stress-energy tensor correlator. In position space, this can be formally written as:
\begin{equation}
\label{graviton self energy}
-i[{}^{\mu\nu}\Sigma^{\rho\sigma}](x;x') = P^{\mu\nu\alpha\beta}(x) P^{\rho\sigma\gamma\delta}(x') \times \partial_\alpha \partial'_\gamma G(x;x') \times \partial_\beta \partial'_\delta G(x;x')
\end{equation}
where $G(x;x')$ is the scalar propagator and the tensor projector is defined as:
\begin{equation}
\label{eq:TensorProjector}
P^{\mu \nu \rho \sigma}=-\frac{\kappa}{2}\left(g^{\mu\rho} g^{\nu\sigma} + g^{\mu\sigma} g^{\nu\rho}-g^{\mu \nu} g^{\rho \sigma}\right)
\end{equation}
The one-loop correction to the quantum effective action is then:
\begin{equation}
\label{oneloopcorr_gen}
S^{(2)}_{corr} = \int d^D x \int d^D x' \, h_{\mu\nu}(x)[{}^{\mu\nu}\Sigma^{\rho\sigma}](x;x') 
h_{\alpha\beta}(x')
\end{equation}
This expression is valid for any arbitrary background. There is also a 4-point diagram, but its contribution vanishes for a massless minimally coupled scalar~\cite{Woodard:Scalar_contribution}. In the following sections, we evaluate this explicitly for flat and de Sitter spacetimes.

\section{One loop graviton self energy in flat spacetime}
\label{sec:oneloop-Minkowski}

We first consider the theory in $D$-dimensional Minkowski space. The action is:
\begin{equation}
S=\frac{1}{2 \kappa^2} \int d^D x \sqrt{-g} R-\frac{1}{2} \int d^D x \sqrt{-g} g^{\mu \nu} \partial_\mu \phi \partial_\nu \phi
\end{equation}
\subsection{Momentum-space approach}

We begin by briefly recapitulating the momentum-space calculation performed in our earlier work~\cite{Mandal:2024kic}. The propagator for a massless minimally coupled scalar field in flat spacetime is given in momentum space by:
\begin{equation}
G(x,x') = \int \frac{d^D k}{(2\pi)^D} \frac{ie^{-ik(x-x')}}{k^2 - i\epsilon}
\end{equation}
To evaluate the self-energy corrections, we compute the integral corresponding to the Feynman diagram in Fig. \ref{fig:scalar corr feynman}:
\begin{equation}
-i[{}^{\mu\nu}\Sigma^{\alpha\beta}](p) = -\int \frac{d^D k}{(2\pi)^D} \frac{1}{(p+k)^2 k^2} P^{\mu\nu\rho\sigma} P^{\alpha\beta\gamma\delta}(p+k)_\rho k_\sigma (p+k)_\gamma k_\delta
\end{equation}
Evaluating this integral using dimensional regularization with $\epsilon = 4-D$, we isolate the divergent contributions. In the limit $D\to4$, the one-loop correction to the effective action takes the local form~\cite{tHooft:1974toh, Capper:1974ed,Deser:1974cz}:
\begin{equation}\label{oneloopcorr}
S^{(2)}_{corr} = -\frac{\alpha}{4\epsilon} 
\int d^4 x \left[ 8\kappa^2 h^{\alpha\beta} (\Box)^2 h_{\alpha\beta} + 4\kappa^2 h (\Box)^2 h \right]
\end{equation}
where the coupling constant $\alpha$ is defined as:
\begin{equation}
\alpha = \frac{1}{120(4\pi)^2}
\end{equation}
To interpret this divergence physically, we compare it to the general quadratic gravity action:
\begin{equation}
\begin{aligned}
S_{\mathrm{quad}}&=\int d^4 x \sqrt{-g}\left[f R_{\mu \nu \rho \sigma} R^{\mu \nu \rho \sigma}+b R_{\mu \nu} R^{\mu \nu}+c R^2\right]^{(2)} \\ 
%%%%
&= \int d^4 x \frac{1}{4}\left[(4 f +b) h^{\mu \nu}(\Box)^2 h_{\mu \nu}+(c-f) h(\Box)^2 h\right]
\end{aligned}
\end{equation}
where $f, b,$ and $c$ are coupling constants. We have employed the Lorenz gauge ($\partial^\mu h_{\mu\nu} = \frac{1}{2}\partial_\nu h$) to obtain the second expression. For Gauss-Bonnet gravity, the couplings satisfy $b = -4f$ and $c = f$. This combination renders the kinetic terms in the quadratic action zero, confirming that the GB term is topological and possesses no propagating degrees of freedom in 4D.

Comparing Eq. \eqref{oneloopcorr} with the general quadratic action, we identify the coefficients:
\begin{equation}
4 \, f + b = -\frac{8\alpha}{\epsilon} , \quad 
c- f = -\frac{4\alpha}{\epsilon}
\end{equation}
To extract the Gauss-Bonnet contribution, we redefine the constants as:
\begin{equation}
f = -\frac{\alpha}{\epsilon} \Bar{f}, \quad b = -\frac{\alpha}{\epsilon} \Bar{b}, \quad c = -\frac{\alpha}{\epsilon} \Bar{c}
\end{equation}
The matching conditions become $4\Bar{f} + \Bar{b} = 8$ and $\Bar{c} - \Bar{f} = 4$. Expressing $\Bar{b}$ and $\Bar{c}$ in terms of $\Bar{f}$, the quantum correction can be rewritten as:
\begin{equation}
\begin{aligned}S_{c o r r}^{(2)} & =-\frac{\alpha}{\epsilon} \bar{f} \int d^4 x \sqrt{-g}\left[R_{\mu \nu \rho \sigma} R^{\mu \nu \rho \sigma}-4 R_{\mu \nu} R^{\mu \nu}+R^2\right] \\ 
%%%
& \quad -\frac{\alpha}{\epsilon} \int d^4 x \sqrt{-g}\left[8 R_{\mu \nu} R^{\mu \nu}+4 R^2\right] \, .
\end{aligned}
\end{equation}
The first integral is the Gauss-Bonnet term. Crucially, its prefactor is proportional to $1/\epsilon = 1/(4-D)$. If the coupling $\alpha$ is rescaled as $\alpha \to \alpha/(D-4)$, the divergence cancels the topological vanishing factor of the GB term, resulting in a finite, anomaly-like quantum correction in 4D. The second term ($8R_{\mu\nu}^2 + 4R^2$) represents the necessary counterterm to cancel the remaining divergence.

We note that because the Gauss-Bonnet density $\mathcal{G}$ is a total derivative in $D=4$, it vanishes identically up to boundary terms. Consequently, the term proportional to $\mathcal{G}/(D-4)$ yields a strictly finite quantum correction to the effective action. It does not cancel UV divergences, which are instead regulated by standard higher-derivative counterterms, but acts as a finite quantum modification analogous to the trace anomaly.

This recovers the primary result obtained in our earlier work~\cite{Mandal:2024kic}. However, this derivation relies on Fourier space integration, which is only valid where global translation invariance exists. To extend this analysis to curved backgrounds—specifically de Sitter spacetime, we cannot rely on momentum space variables. We must instead rederive these results using real-space (coordinate space) techniques to ensure the method is robust before applying it to curved geometry. We proceed with this real-space formulation in the following subsection.

\subsection{Real-space approach}

In $D$-dimensional flat spacetime, the Schwinger–Keldysh scalar propagator is~\cite{Ford:2004wc}:
\begin{equation}
G(x,x') = \frac{\Gamma(\frac{D}{2} - 1)}{4\pi^{\frac{D}{2}}} \left(\frac{1}{\Delta x^2}\right)^{\frac{D}{2} - 1}
\end{equation}
where $\Delta x^2 \equiv ||\mathbf{x} - \mathbf{x}'||^2 - (t-t')^2$. The one-loop graviton self-energy tensor in real space can be decomposed as:
\begin{equation}
\label{126}
\begin{aligned}
-i[{}^{\mu\nu}\Sigma^{\rho\sigma}]_{\mathrm{flat}} = & \frac{\kappa^2 \Gamma^2\left(\frac{D}{2}\right)}{16 \pi^D}\left
\{ \eta^{\mu(\rho} \eta^{\sigma) \nu} \left[-\frac{2}{\Delta x^{2 D}}\right]+\Delta x^{(\mu} \eta^{\nu)(\rho} \Delta x^{\sigma)} \left[\frac{4 D}{\Delta x^{2 D+2}}\right] \right. \\
%%%%
& +\Delta x^\mu \Delta x^\nu \Delta x^\rho \Delta x^\sigma \left[-\frac{2 D^2}{\Delta x^{2 D+4}}\right]+\eta^{\mu \nu} \eta^{\rho \sigma} \left[-\frac{1}{2} \frac{\left(D^2-D-4\right)}{\Delta x^{2 D}}\right] \\
%%%%
& \left.+\left[\eta^{\mu \nu} \Delta x^\rho \Delta x^\sigma+\Delta x^\mu \Delta x^\nu \eta^{\rho \sigma}\right] \left[\frac{D(D-2)}{\Delta x^{2 D+2}}\right]\right\}
\end{aligned}
\end{equation}
We can express this tensor structure in terms of projection operators $\Pi^{\mu\nu} = \partial^\mu \partial^\nu - \eta^{\mu\nu} \partial^2$:
\begin{equation}
-i[{}^{\mu\nu}\Sigma^{\rho\sigma}]_{\mathrm{flat}} =\Pi^{\mu \nu} \Pi^{\rho \sigma}G\left(\Delta x^2\right)+\Pi^{\mu(\rho} \Pi^{\sigma) \nu} F\left(\Delta x^2\right)
\end{equation}
where the scalar functions are found to be:
\begin{align}
F\left(\Delta x^2\right) &= -\frac{\kappa^2 \Gamma^2\left(\frac{D}{2}\right)}{16 \pi^D} \frac{1}{4(D-2)^2(D-1)(D+1)}\left(\frac{1}{\Delta x^2}\right)^{D-2}\\ 
%%%%
G\left(\Delta x^2\right) &= -\frac{\kappa^2 \Gamma^2\left(\frac{D}{2}\right)}{16 \pi^D} \frac{(D-2)^2 (D+1)-2}{8(D-1)^2(D-2)^2(D+1)}\left(\frac{1}{\Delta x^2}\right)^{D-2}
\end{align}
The corrected action in terms of these form factors is:
\begin{equation}
\begin{aligned}
S_{corr} =& \int d^4x \, h_{\mu\nu} \left[{ }^{\mu \nu} \Sigma^{\rho \sigma}\right]_{\mathrm{flat}} h_{\rho\sigma} \\
%%%
=& F (h_{\mu\nu}(\Box)^2 h^{\mu\nu}) + G(h(\Box)^2 h) + (F + G)(\partial^\mu \partial^\nu h_{\mu\nu} \partial^\rho \partial^\sigma h_{\rho\sigma})\\ 
%%%%
&+2F (\partial^\mu h_{\mu\nu} \Box \partial_\rho h^{\nu\rho}) - 2G(\partial^\mu\partial^\nu h_{\mu\nu} \Box h)
\end{aligned}
\end{equation}

While the underlying UV divergences are strictly minimized by the Weyl-squared ($C^2$) and Ricci-scalar-squared ($R^2$) invariants, we project these onto an alternative basis containing the Gauss-Bonnet density to explicitly isolate the finite scaling terms. Linearizing the generic quadratic gravity action with respect to the Minkowski background yields:
\begin{equation}
\begin{aligned}
\sqrt{-g} &(f R_{\mu \nu \rho \sigma}^2+b R_{\mu \nu}^2+c R^2) \\
%%%
& =\left(f+\frac{b}{4}\right) h^{\mu \nu}(\Box)^2 h_{\mu \nu}+\left(\frac{b}{4}+c\right) h(\Box)^2 h + \left( f + \frac{b}{2} + c \right) \partial_\mu \partial_\nu h^{\mu\nu} \partial_\rho \partial_\sigma h^{\rho\sigma}\\
&+ \left( 2f + \frac{b}{2} \right) \partial_\sigma h^{\sigma\mu} \Box \partial^\rho h_{\rho\mu} - \left( \frac{b}{2} + 2c \right) \partial_\mu\partial_\nu h^{\mu\nu} \Box h
\end{aligned}
\end{equation}
Comparing the coefficients of the box-squared terms:
\begin{equation}
F\left(\Delta x^2\right) = f + \frac{b}{4}; 
\quad G\left(\Delta x^2\right) = \frac{b}{4} + c
\end{equation}
The term $\left(1/\Delta x^2\right)^{D-2}$ contains the logarithmic divergence. To isolate the pole, we use the differential identity:
\begin{equation}
\left(\frac{1}{\Delta x^2}\right)^{D-2} = \frac{\Box}{2(D-3)(D-4)}\left(\frac{1}{\Delta x^2}\right)^{D-3}
\end{equation}
Since $\Box (1/\Delta x^2)^{D-3} \sim \delta^D(x-x')$, this confirms the divergence is local. Extracting the pole at $D=4$:
\begin{align}
4 \, f + b &= 4F = \frac{8\alpha}{D-4} \\ 
%%%
c - f &= G - F = \frac{4\alpha}{D-4}
\end{align}
Here we have identified the prefactor $\alpha$ consistently with the momentum space result:
\begin{equation}
\alpha = -\frac{\kappa^2}{(4\pi)^4 120} \Box \left(\frac{1}{\Delta x^2}\right)_{\text{finite}}
\end{equation}
Thus, the real-space calculation reproduces the exact same $1/(D-4)$ pole structure, validating the method for application to curved spacetime.

\section{Scalar contribution to Graviton self-energy in de Sitter}
\label{sec:oneloop-scalar-deSitter}

Having established the validity of the real-space method in flat spacetime, we now extend our analysis to a curved background. We emphasize that, due to the local nature of ultraviolet divergences, the structural coefficients of the counterterms are strictly independent of the background geometry. The primary motivation for extending our real-space analysis to an explicit de Sitter background is therefore twofold: first, to track the background-dependent finite field corrections proportional to the Hubble parameter $H$, which dictate early-universe phenomenology; and second, to demonstrate the background-robustness of our coordinate-space regularization scheme.

We consider $D$-dimensional de Sitter ($dS_D$) spacetime, which serves as the maximally symmetric solution to the vacuum Einstein equations with a positive cosmological constant. For the specific details regarding the embedding of de Sitter space, the Poincar\'e patch coordinates used in this calculation, and the properties of the propagator, we refer the reader to the Appendix.

\subsection{Graviton self-energy correction in de Sitter}

In curved spacetime, the definition of the self-energy requires handling the breaking of translation invariance. Using the covariant real-space formalism, the divergent part of the one-loop graviton self-energy, $\left[{ }_{\mu \nu} \Sigma_{\rho \sigma}^{\text {div}}\right](x;x')$, can be expressed in terms of bitensors (functions of two points $x$ and $x'$). Following the analysis in Ref.~\cite{kamperman2025one}, the divergent contribution is given by:
\begin{equation}
\begin{aligned}
& -i\left[{ }_{\mu \nu} \Sigma_{\rho \sigma}^{\text {div}}\right]
\left(x ; x^{\prime}\right) \\
&= \frac{ \kappa^2}{16 \pi^2}\left\{\frac{1}{8(D+1)(D-1)} 
\mathcal{F}_{\mu \nu \rho \sigma}+\frac{1}{2}\left(\frac{D-2}{4(D-1)}\right)^2 \mathcal{G}_{\mu \nu \rho \sigma}\right\} \frac{H^{D-4}}{D-4} \frac{i \delta^D\left(x-x^{\prime}\right)}{\sqrt{-g}}
\end{aligned}
\end{equation}
Here, $\mathcal{F}_{\mu \nu \rho \sigma}$ and $\mathcal{G}_{\mu \nu \rho \sigma}$ are fourth-rank tensor differential operators constructed from the background metric and covariant derivatives. These structures encapsulate the geometric complexity of the loop integral in a curved background. They are explicitly defined as:
\begin{equation}
\begin{aligned}
\mathcal{G}_{\mu \nu \rho \sigma}= & 
2 \nabla_\mu \nabla_\nu \nabla_\rho^{\prime} \nabla_\sigma^{\prime}+2 g_{\mu \nu} g_{\rho \sigma} \square \square^{\prime}-2\left(g_{\mu \nu} \nabla_\rho^{\prime} \nabla_\sigma^{\prime}+g_{\rho \sigma} \nabla_\mu \nabla_\nu\right) \square \\ 
%%%%
& +H^2\left[(D-1)(D-2)\left(g_{\mu \nu} \nabla_\rho^{\prime} \nabla_\sigma^{\prime}+g_{\rho \sigma} \nabla_\mu \nabla_\nu\right)-2 D(D-1) g_{\mu)(\rho} \nabla_{\sigma)} \nabla_{(\nu}\right. \\ 
& \left.+(D-4)(D-1)^2 g_{\mu \nu} g_{\rho \sigma} \square+D(D-1) g_{\mu(\rho} g_{\sigma) \nu} \square\right] \\ 
%%%%
& +H^4 \frac{(D-1)^2}{4}\left[\left(D^2-8 D+8\right) g_{\mu \nu} g_{\rho \sigma}+2 D^2 g_{\mu(\rho} g_{\sigma) \nu}\right],
\end{aligned}
\end{equation}
and
\begin{equation}
\begin{aligned}
\mathcal{F}_{\mu \nu \rho \sigma}= & 
\frac{D-2}{D-1} \nabla_\mu \nabla_\nu \nabla_\rho^\prime  \nabla_\sigma^{\prime}+\frac{1}{D-1}\left(g_{\mu \nu} \nabla_\rho^{\prime} \nabla_\sigma^{\prime}+g_{\rho \sigma} \nabla_\mu \nabla_\nu\right) \square-2 g_{\mu)(\rho} \nabla_{\sigma)} \nabla_{(\nu} \square \\ 
%%%
& -\frac{1}{D-1} g_{\mu \nu} g_{\rho \sigma} \square \square^{\prime}+g_{\mu(\rho} g_{\sigma) \nu} \square \square^{\prime} \\ 
%%%
& +H^2\left[(D-2)\left(g_{\mu \nu} \nabla_\rho^{\prime} \nabla_\sigma^{\prime}+g_{\rho \sigma} \nabla_\mu \nabla_\nu\right)-2 D g_{\mu)(\rho} \nabla_{\sigma)} \nabla_{(\nu}\right. \\
%%%
& \left.+(D-3) g_{\mu \nu} g_{\rho \sigma} \square-2(D-1) g_{\mu(\rho} g_{\sigma) \nu} \square\right] \\ 
%%%
& +H^4\left(D(D-1)^2+(D-2)\right)\left[g_{\mu \nu} g_{\rho \sigma}-D g_{\mu(\rho} g_{\sigma) \nu}\right] .
\end{aligned}
\end{equation}
The appearance of terms proportional to $H^2$ and $H^4$ reflects the non-trivial coupling between the matter loop and the background curvature.
To obtain the correction to the effective action, we contract this self-energy with the graviton fluctuations $h_{\mu\nu}$. Integration by parts allows us to move derivatives from the self-energy kernel onto the graviton fields. Evaluating the integral explicitly in $D=4$ dimensions, we obtain:
\begin{equation}\label{self energy pert dS}
\begin{aligned}
S_{corr} = -\int d^Dx d^Dx'& \, h^{\mu\nu}(x) \left[{ }_{\mu \nu} \Sigma_{\rho \sigma}^{\mathrm{div}}\right] (x;x') h^{\rho\sigma}(x') \\
%%%%
= -\frac{\alpha}{(D-4)}\int& d^4x \sqrt{-g} \left[80 H^4 h_{\mu\nu} h^{\mu\nu} - 20 H^4 h^2 - 112 H^2 h^{\mu\nu} \nabla^\rho \nabla_\mu h_{\nu\rho}\right.\\ 
%%%
& + 8 H^2 ( h^{\mu\nu}\nabla_\mu\nabla_\nu h + h \nabla_\mu\nabla_\nu h^{\mu\nu} ) - 32 H^2 h \nabla_\mu \nabla_\nu h^{\mu\nu} \\ 
%%%
& + 14 H^2 h^{\mu\nu} \Box h_{\mu\nu} + 7 H^2 h \Box h + 4 h^{\mu\nu}\nabla_\rho\nabla_\sigma\nabla_\mu\nabla_\nu h^{\rho\sigma}  \\ 
%%%
&\left. - 2h^{\mu\nu} \Box \nabla^\rho \nabla_\nu h_{\mu\rho} - 6 h \Box \nabla_\mu \nabla_\nu h^{\mu\nu} + 3 h (\Box)^2 h + h^{\mu\nu} (\Box)^2 h_{\mu\nu} \right].
\end{aligned}
\end{equation}
This expression represents the raw divergent contribution to the effective action coming from the matter loop in a de Sitter background.

\subsection{Comparison with the Gauss-Bonnet term}

To interpret the physical content of Eq. \eqref{self energy pert dS}, we must compare it with the expansion of the general quadratic curvature action in de Sitter space. Expanding the action 
$$S_{\mathrm{quad}} = \int d^4x \sqrt{-g} (f\, R_{\mu\nu\rho\sigma}^2 + b\, R_{\mu\nu}^2 + c\, R^2)$$ 
to second order in perturbations $h_{\mu\nu}$ around the dS metric yields:
\begin{equation}
\begin{aligned}
S_\mathrm{quad} &= \int d^4x \sqrt{-g} \left[ 4(4f+5b+16c)H^4 h_{\mu\nu}h^{\mu\nu} - (4f+5b+16c)H^4 h^2 \right.\\ 
&+ (3f+2b+5c) H^2 (h^{\mu\nu} \nabla_\mu \nabla_\nu h + h \nabla_\mu \nabla_\nu h^{\mu\nu}) - 2(6f+5b+14c) H^2 h^{\mu\nu} \nabla^\rho \nabla_\mu h_{\rho\nu} \\ 
&+\frac{1}{2} (8f+5b+12c) H^2 h^{\mu\nu} \Box h_{\mu\nu} - \left( f + \frac{b}{4} \right) H^2 h \Box h + \left( f +\frac{b}{2}+c \right) h^{\mu\nu} \nabla_\rho\nabla_\sigma\nabla_\mu\nabla_\nu h^{\rho\sigma} \\ 
&-\left( 2 f + \frac{b}{2} \right) h^{\mu\nu} \Box \nabla^\rho\nabla_\mu h_{\rho\nu} - \left( \frac{b}{2} + 2c \right) h \Box \nabla_\mu\nabla_\nu h^{\mu\nu} \\ 
&\left.+ \left( f + \frac{b}{4} \right) h^{\mu\nu} (\Box)^2 h_{\mu\nu} + \left( \frac{b}{4} + c \right) h (\Box)^2 h \right]
\end{aligned}
\end{equation}
We now equate the coefficients of the independent tensor structures in the quantum corrected action \eqref{self energy pert dS} with those in the general geometric expansion. Focusing on the highest-derivative terms (the $(\Box)^2$ terms), we find the same constraints as in the Minkowski case:
\begin{equation}
f + \frac{b}{4} \propto 1 \quad \text{and} \quad 
\frac{b}{4} + c \propto 3
\end{equation}
Solving the system of linear equations for all coefficients ($H^4$, $H^2$, and derivative terms) consistently yields a solution where the couplings $f, b, c$ align with the Gauss-Bonnet combination ($b = -4 f, c=f$) plus a residual counterterm. Specifically, we obtain:
\begin{equation}
\begin{aligned}
\mathcal{S}_\mathrm{corr} = \frac{\alpha}{(D-4)} \bar{f} & \int d^4x \sqrt{-g} \left[ R_{\mu\nu\rho\sigma}R^{\mu\nu\rho\sigma} - 4 R_{\mu\nu}R^{\mu\nu} + R^2 \right] \\ 
+ \frac{\alpha}{2(D-4)} & \int d^4x \sqrt{-g} \left[ 8R_{\mu\nu}R^{\mu\nu} + 4R^2 \right]
\end{aligned}
\end{equation}
where $\bar{f}$ is a finite constant. This is the key result of this work regarding which we want to discuss the following points: To begin with, the first term is the Gauss-Bonnet invariant. The $1/(D-4)$ pole from the loop integral precisely matches the scaling required by the Glavan-Lin proposal~\cite{Glavan:2019inb} to yield a finite, non-trivial contribution in the $D \to 4$ limit. Furthermore, the second term represents the divergent counterterm required to renormalize the theory. Notably, the structure of this counterterm ($8R_{\mu\nu}^2 + 4R^2$) is identical to the one found in the Minkowski analysis (up to overall normalization factors consistent with the dS volume element)~\cite{Nenmeli:2021orl}. Finally, this confirms that the renormalization of the one-loop matter sector leads to a background-independent result: a finite Gauss-Bonnet quantum correction and a specific Ricci-squared counterterm. This counterterm is similar to the term usually added to the self-energy to remove local divergent terms~\cite{kamperman2025one}.

Finally, it is important to address the uniqueness and locality of this effective action. In perturbative quantum field theory, off-shell quadratic curvature counterterms are not strictly unique under metric field redefinitions of the form $g_{\mu\nu} \to g_{\mu\nu} + \delta g_{\mu\nu}$, where $\delta g_{\mu\nu}$ depends linearly on the Ricci tensor or Ricci scalar. While such redefinitions leave the S-matrix and on-shell physical observables invariant, they can shift off-shell structures between Riemann-squared, Ricci-squared, and scalar-squared terms, or modify them by total derivatives. Our specific projection onto the Gauss-Bonnet basis is an algebraic choice designed to demonstrate how loop-level poles interact with topological invariants to mirror phenomenological 4D Einstein-Gauss-Bonnet models. Additionally, while the specific scaling mechanism discussed here arises from local ultraviolet poles canceling against local vanishing factors, a complete quantum effective action inevitably incorporates non-local finite contributions. In momentum space, these correspond to logarithmic form factors $\sim \ln(\Box/\mu^2)$, which manifest as non-local logarithmic tail functions accompanying the local divergence structures in real space.

\section{Photon contribution to graviton self-energy in de Sitter} 
\label{sec:oneloop-photon-deSitter}

We now turn our attention to the contribution of electromagnetic (photon) loops to the graviton self-energy. This calculation presents distinct challenges compared to the scalar case. While the massless scalar field is conformally coupled to gravity only for a specific value of the non-minimal coupling $\xi$, the Maxwell field is classically conformally invariant in $D=4$. However, since we employ dimensional regularization ($D \neq 4$) to handle divergences, this conformal symmetry is explicitly broken during the intermediate stages of the calculation. This breaking is non-trivial; terms proportional to $(D-4)$ interact with the $1/(D-4)$ divergence to yield finite, anomalous contributions to the effective action, a phenomenon intimately related to the trace anomaly~\cite{Duff:1993wm}. 

The starting point is the Einstein-Maxwell Lagrangian in $D$-dimensions:
\begin{equation}
\mathcal{L} = \frac{1}{2\kappa^2} \sqrt{-g} 
\left[ R  - (D-1)(D-2)H^2 \right] - \frac{1}{4} \sqrt{-g} F_{\mu\nu} F^{\mu\nu}
\end{equation}
To quantize the theory and define the propagators, we must break the gauge invariance. The choice of gauge is particularly important in de Sitter space to ensure the resulting propagators respect the symmetries of the background maximally. We employ the \emph{exact gauge} (or average gauge) introduced in Ref.~\cite{woodard2006sitter}. This gauge choice is designed to simplify the photon propagator structure, effectively reducing it to a form involving massless scalar propagators.

The gauge-fixing term for the electromagnetic sector is chosen as:
\begin{equation}
\mathcal{L}^\mathrm{EM}_\mathrm{GF} = 
-\frac{1}{2} a^{D-4} \left( \eta^{\mu\nu} A_{\mu,\nu} - (D-4) H a A_0 \right)^2
\end{equation}
where $a \equiv a(\eta) = -\frac{1}{H\eta}$ is the conformal scale factor of the de Sitter background (see Appendix), and the indices in this gauge-fixing term are contracted using the flat Minkowski metric $\eta^{\mu\nu}$. Note the explicit appearance of $(D-4)$. For $D = 4$, this reduces to the standard Lorenz gauge $\nabla_\mu A^\mu = 0$ (in conformally flat coordinates). However, for $D \neq 4$, this specific modification ensures that the longitudinal modes decouple cleanly, allowing the photon propagator to be expressed simply in terms of the scalar propagator.

We emphasize that while off-shell self-energies and intermediate bitensor structures generally exhibit gauge dependence, standard quantum field theoretic Ward and BRST identities guarantee that the ultraviolet divergent counterterms extracted from the loop integral correspond to strictly gauge-independent geometric invariants. The explicit $(D-4)$ dimension dependence in our gauge-fixing condition is engineered solely to facilitate longitudinal decoupling away from four dimensions. Because this modifying term vanishes identically in the $D \to 4$ limit, it does not introduce spurious divergences or affect the gauge invariance of the extracted physical counterterms, ensuring that the resulting Weyl-squared invariant and its associated anomaly coefficients remain gauge-independent.

Similarly, for the graviton sector, we use a de Donder-type gauge adapted to the de Sitter geometry~\cite{Tsamis:1992xa}:
\begin{equation}
\mathcal{L}^\mathrm{GR}_\mathrm{GF} = -\frac{1}{2} a^{D-2} \eta^{\mu \nu} 
\Lambda_\mu \Lambda_\nu \quad, \quad \Lambda_\mu \equiv \eta^{\rho \sigma}\left(\partial_\sigma h_{\mu \rho}-\frac{1}{2} \partial_\mu h_{\rho \sigma}+(D-2) H a \delta_\sigma^0 h_{\mu \rho}\right)
\end{equation}
\begin{figure}[t]
    \centering
    \includegraphics[width=0.5\linewidth]{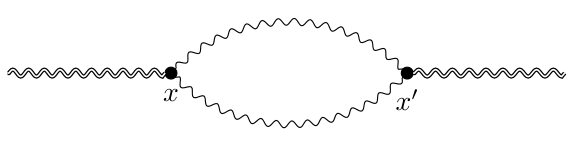}
    \caption{One-loop Feynman diagram describing the self-energy correction to the graviton propagator due to a photon.}
    \label{fig:photon corr feynman}
\end{figure}

The relevant Feynman diagram is shown in Figure \ref{fig:photon corr feynman}. The calculation of this diagram involves contracting the vertex factors derived from the interaction Lagrangian $-\frac{1}{4}\sqrt{-g}F^2$ with the photon propagators. The resulting self-energy tensor can be expressed in terms of bitensors (functions of $x$ and $x'$). Following the methodology in Ref.~\cite{Wang:2015eaa}, the self-energy is written as a differential operator acting on a scalar structure function:
\begin{equation}
-i[{}^{\mu\nu}\Sigma^{\rho\sigma}] \, (x',x) = \tilde{\mathcal{F}}^{\mu\nu\rho\sigma}[F_2(x',x)]
\end{equation}
where $\tilde{\mathcal{F}}^{\mu\nu\rho\sigma}$ is a fourth-order differential operator constructed from the background metric and derivatives:
\begin{equation}
\tilde{\mathcal{F}}^{\mu\nu\rho\sigma} = \mathcal{P}^{\mu\nu}_{\alpha\beta\gamma\delta} \mathcal{P}^{\rho\sigma}_{\kappa\lambda\theta\phi} g^{\alpha\kappa} g^{\beta\lambda} g^{\gamma\theta} g^{\delta\phi}
\end{equation}
The tensor projector $\mathcal{P}$ is defined in Ref.~\cite{Woodard:Scalar_contribution}. The core physical content of the loop is contained in the structure function $F_2(x;x')$. The divergent part of this function, which requires renormalization, is~\cite{Wang:2015eaa}:
\begin{equation}
F_2^\mathrm{div} = -\frac{\kappa^2 \mu^{D-4} \Gamma\left(\frac{D}{2}\right)}{128 \pi^{\frac{D}{2}}} \frac{D(D-2) i \delta^D\left(x-x^{\prime}\right)}{(D-4)(D-3)^2(D+1)} \approx -i \frac{\kappa^2}{16 \pi^2} \frac{1}{5} \frac{\delta^D(x-x')}{(D-4)}
\end{equation}
We observe the characteristic $1/(D-4)$ pole. By substituting this back into the effective action formula and integrating by parts to move derivatives onto the external graviton fields $h_{\mu\nu}$, we obtain the one-loop correction to the effective action:
\begin{equation}\label{EM self energy pert dS}
\begin{aligned}
S_{corr} =& -\int d^Dx d^Dx' , h_{\mu\nu}(x) \left[{ }^{\mu \nu} \Sigma^{\rho \sigma}\right]^{\mathrm{div}} h_{\rho\sigma}(x') \\ 
=& -\frac{\beta}{(D-4)}\int d^4x \sqrt{-g} \left[16 H^4 h_{\mu\nu} h^{\mu\nu} - 4 H^4 h^2 - 4 H^2 ( h^{\mu\nu}\nabla_\mu\nabla_\nu h + h \nabla_\mu\nabla_\nu h^{\mu\nu} )\right.\\ 
&+ 8 H^2 h^{\mu\nu} \nabla_\rho \nabla_\mu h_\nu^\rho- 6 H^2 h^{\mu\nu} \Box h_{\mu\nu} + 3 H^2 h \Box h - 2 h^{\mu\nu}\nabla_\rho\nabla_\sigma\nabla_\mu\nabla_\nu h^{\rho\sigma} \\
&\left. + 6 h^{\mu\nu} \Box \nabla_\rho \nabla_\nu h_\mu^\rho - 2 h \Box \nabla_\mu \nabla_\nu h^{\mu\nu} +  h \Box^2 h - 3 h^{\mu\nu} \Box^2 h_{\mu\nu} \right]
\end{aligned}
\end{equation}
where $\beta$ encapsulates the coupling constants. This expression represents the raw UV divergence of the theory. To interpret this physically, like the scalar field case, we compare the tensor structure of $S_{corr}$ with the expansion of the general quadratic curvature action $S_{quad} = \int \sqrt{-g}(f R_{\mu\nu\rho\sigma}^2 + b R_{\mu\nu}^2 + c R^2)$.

Matching the coefficients of the highest derivative terms (specifically the $\Box^2$ terms), we identify the constraints on the couplings:
\begin{equation}
f + \frac{b}{4} = -3 \frac{\beta}{(D-4)}, \quad 
c - f =\frac{\beta}{(D-4)}
\end{equation}
Interestingly, just as in the scalar case, the solution to these constraints yields the Gauss-Bonnet combination plus a specific counter-term. The renormalized action can be structured as:
\begin{equation}
\begin{aligned}
\mathcal{S}_\mathrm{corr} =& \frac{\beta}{(D-4)} \bar{f} \int d^4x \sqrt{-g} \left[ R_{\mu\nu\rho\sigma}R^{\mu\nu\rho\sigma} - 4 R_{\mu\nu}R^{\mu\nu} + R^2 \right] \\ 
+& \frac{4\beta}{(D-4)} \int d^4x \sqrt{-g} \left[R^2 -3R_{\mu\nu}R^{\mu\nu}  \right]
\end{aligned}
\end{equation}
The first term is the familiar topological Gauss-Bonnet invariant, rendered into a finite, local quantum correction by the $1/(D-4)$ pole. The second term is the counter-term required to cancel the divergence. This counter-term is equivalent to the Weyl-squared (or conformal gravity) term. Using the definition of the Weyl tensor $C_{\mu\nu\rho\sigma}$, the counter-term action can be rewritten as:
\begin{equation}
\begin{aligned}
\mathcal{S}_\mathrm{ct} &= \int d^4 x \sqrt{-g} \, c_2 \,  C_{\mu\nu\rho\sigma} C^{\mu\nu\rho\sigma} \\
&= \int d^4 x \sqrt{-g} \, c_2 \,  \left[ \mathcal{G} + \frac{2}{3}\left( 3R_{\mu\nu}R^{\mu\nu} - R^2 \right)  \right]
\end{aligned}
\end{equation}
where the coefficient $c_2$ is:
\begin{equation}
c_2=\frac{\mu^{D-4} \Gamma\left(\frac{D}{2}\right)}{256 \pi^{\frac{D}{2}}} \frac{D(D-2)}{(D-4)(D-3)^2(D+1)} .
\end{equation}
This confirms that the one-loop photon self-energy induces a Gauss-Bonnet term and a Weyl-squared counter-term, a structure directly linked to the conformal anomaly of the electromagnetic field in curved spacetime~\cite{Duff:1977ay}.

\section{Implications for the Early Universe}
\label{sec:Early-Universe}

The results derived in the previous sections have important implications for the dynamics of the early universe. We have demonstrated that the self-energy corrections from matter loops (scalars and photons) in a curved background naturally generate a finite Gauss-Bonnet quantum correction in the effective action. This occurs because the $1/(D-4)$ ultraviolet pole of the loop integral multiplies the topological Gauss-Bonnet density, yielding a finite residual upon taking the four-dimensional limit. This suggests that in the high-curvature regime of the early universe --- such as the epoch immediately following the Big Bang --- the gravitational dynamics deviate significantly from standard General Relativity. 

Specifically, in a radiation-dominated universe (filled with the highly quantum EM fields discussed above), the modified Friedmann equation in 4D Einstein-Gauss-Bonnet gravity typically takes the form $H^2 + \alpha H^4 \sim \rho_{rad}$. In the ultraviolet limit where the Gauss-Bonnet term dominates ($H^2 \ll \alpha H^4$), and assuming standard radiation scaling $\rho_{rad} \propto a^{-4}$, the expansion dynamics scale as $H \propto a^{-1}$, leading to a linear expansion law $a(t) \propto t$~\cite{Fernandes:2022zrq}. While this represents a \emph{linear coasting universe} that resolves the initial singularity more effectively than the standard $a(t) \propto t^{1/2}$ radiation solution, it does not, by itself, yield the exponential expansion ($\ddot{a} > 0$) required to solve the horizon and flatness problems~\cite{Starobinsky:1980te}.

However, our renormalization procedure reveals a second, critical contribution: the counterterms required to cancel the true ultraviolet divergences strictly consist of quadratic curvature invariants, specifically the Weyl-squared ($C_{\mu\nu\rho\sigma}C^{\mu\nu\rho\sigma}$) and Ricci-scalar-squared ($R^2$) terms. Since, the Weyl tensor vanishes identically on conformally flat Friedmann-Lema\^itre-Robertson-Walker (FLRW) backgrounds, only the $R^2$ counterterm impacts the homogeneous cosmological dynamics. It is well-established that the presence of an $R^2$ term in the action leads to Starobinsky inflation, where the extra scalar degree of freedom (the scalaron) drives an accelerated inflationary phase with a natural graceful exit~\cite{Nenmeli:2021orl,Das:2022hjp}.

Thus, we propose that the mechanism for early universe acceleration arises not solely from the Gauss-Bonnet term, but from the full semi-classical effective action. The matter loops simultaneously activate the Glavan-Lin Gauss-Bonnet dynamics (which may soften the singularity) and the Starobinsky-like dynamics (which drive inflation). This framework offers a unified origin for modified gravity in the early universe: the same loop corrections that yield a finite 4D Gauss-Bonnet term also provide the necessary higher-derivative terms for inflation, without the need to introduce ad-hoc scalar fields.

\section{Discussion}
\label{sec:discussion}

In this work, we have demonstrated that the one-loop self-energy correction to the graviton, driven by scalar and photon loops in a curved background, naturally generates higher-derivative terms in the effective action. Specifically, our dimensional regularization procedure reveals that while standard ultraviolet divergences must be renormalized by the Weyl-squared and Ricci-scalar-squared ($R^2$) counterterms, the interaction of the $1/(D-4)$ UV pole with the topological Gauss-Bonnet invariant yields a finite quantum correction. This finite loop-level modification effectively mirrors the singular scaling proposed in 4D Einstein-Gauss-Bonnet gravity. While the complete quantum effective action inevitably incorporates non-local form factors (such as logarithmic tail functions $\sim \ln(\Box/\mu^2)$ accompanying the local terms), this local finite correction provides a concrete field-theoretic motivation for exploring higher-derivative modifications. A critical open question is how such loop-level modifications to General Relativity can be tested or constrained empirically. 

In the radiation-dominated epoch of the early universe, primordial gravitational waves do not propagate through a true vacuum, but rather through a dense, thermal bath of photons and relativistic particles. Within our framework, the propagation of a graviton through this plasma is modified by one-loop vacuum polarization effects, where the graviton self-energy $[{}_{\mu\nu}\Sigma_{\rho\sigma}]$ receives finite-temperature corrections from the photon bath. 

By analytically continuing the photon structure functions derived in zero-temperature formalisms to finite temperature $T$ (e.g., using the Matsubara formalism), the graviton propagator develops a modified, temperature-dependent dispersion relation~\cite{Rebhan:1990yr,Brandt:1993bj,Brandt:1998hd}:
$$\omega^2 = k^2 + \Sigma_{TT}(\omega, k, T)$$
where $\Sigma_{TT}$ is the transverse-traceless projection of the finite-temperature self-energy~\cite{Brandenberger:1984cz}. To leading order, the thermal photon gas contributes a term proportional to the energy density of the bath, $\Sigma_{TT} \propto G T^4$, where $G$ is Newton's constant. This imparts an effective refractive index to the cosmological spacetime, introducing a frequency-dependent phase velocity (dispersion) for primordial gravitational waves. Furthermore, the imaginary part of this thermal self-energy, $\text{Im}[\Sigma_{TT}]$, can provide a rigorous microscopic derivation of the collisionless damping of tensor perturbations as they interact with the primordial plasma.

Near a black hole horizon or a neutron star, the background gravitational and electromagnetic fields are intense. In standard geometric optics, gravitational waves (GWs) simply follow null geodesics in the curved background. In quantum electrodynamics, Delbr\"uck scattering is the deflection of a photon by the Coulomb field of a nucleus, mediated by a virtual electron-positron loop~\cite{Drummond:1979pp,Jana:2021lqe,Johnson:2023skw}. In our framework, a graviton propagating near a black hole can scatter off the background curvature mediated by a virtual photon loop. This 1-loop effect modifies the classical scattering cross-section of GWs by massive bodies. As shown, integrating out the photon loops yields an effective action containing higher-derivative corrections, including the finite Gauss-Bonnet contribution and the Weyl-squared terms. Near a strong gravity source, these terms cease to be negligible. They modify the effective potential the GW experiences, potentially leading to distinct signatures in the quasi-normal mode spectrum (ringdown) that differ from pure classical General Relativity.

Traditionally, the search for higher-derivative gravity signatures has focused on cosmological observables, such as the primordial power spectrum, or astrophysical phenomena, such as black hole ringdown. However, recent theoretical developments in the quantum detection of gravitational waves suggest a novel, complementary approach. As recently shown in Ref.~\cite{Hari:2026qyt}, the interaction between a gravitational wave and an electromagnetic field in an interferometer can be rigorously formulated as a scattering process between a quantized electromagnetic field and a coherent state of gravitons. In this framework, the standard geometric time delay emerges from the phase shifts induced by the absorption and emission of gravitons, leading to measurable changes in two-photon coincidence rates via Hong-Ou-Mandel (HOM) interference.

The generation of the finite Gauss-Bonnet correction and Weyl-squared terms implies that the low-energy effective field theory of gravity deviates from the pure Einstein-Hilbert action. Consequently, both the graviton propagator and the photon-graviton interaction vertex receive quantum corrections. When computing the scattering amplitude between the incoming photon states and the coherent graviton state, these corrections will introduce momentum-dependent modifications to the tree-level amplitudes derived in Ref.~\cite{Hari:2026qyt}. We are interested in investigating how large these corrections are and their observational implications in future work.

\acknowledgments
The authors gratefully acknowledge S. Mandal for his collaboration and valuable contributions during the early stages of this work. The authors thank P. George Christopher, K. Hari, and T. Parvez for comments on the earlier draft. The work is supported by the ANRF
Advanced Research grant (ANRF/ARG/2025/001514/PS).
This work is part of the Undergraduate project of AK.

\vspace{-0.5cm}
\section*{Author Contributions}
A.K. did the calculations. S.S. motivated the idea and supervised. Both authors wrote the manuscript.

\vspace{-0.5cm}
\section*{Conflicts of Interest}

The authors declare no conflicts of interest.

\vspace{-0.5cm}
\section*{Data Availability Statement}

Data sharing not applicable to this article as no datasets were generated or analysed during the current study.

\appendix
\section{de Sitter space}
The $d$-dimensional de Sitter space, denoted as $dS_d$, is the maximally symmetric solution to the vacuum Einstein field equations with a positive cosmological constant $\Lambda$. It plays a pivotal role in modern cosmology, serving as the standard model for both the inflationary epoch of the early universe and the current era of accelerated expansion. As detailed in the seminal work by Hawking and Ellis~\cite{Hawking:1973uf}, $dS_d$ has the topology $\mathbb{R}\times S^{d-1}$ and can be visualized as a hyperboloid embedded in a $(d+1)$-dimensional Minkowski spacetime.

The embedding is defined by the hypersurface equation:
\begin{equation}
\label{d+1}
\begin{aligned}
\eta_{MN} X^M X^N &= \frac{1}{H^2} \quad (M, N,\cdots = 0,\dots,d) \\ 
%%%
\eta_{MN} &= \text{diag (-1,1,\dots,1)}
\end{aligned}
\end{equation}
where the ambient metric is given by:
\begin{equation}
ds^2 = \eta_{MN} dX^M dX^N
\end{equation}
Here, $H$ is the Hubble constant, related to the cosmological constant by $\Lambda = \frac{(d-1)(d-2)}{2}H^2$. The isometry group of this space is the Lorentz group $SO(1, d)$ of the embedding space, which acts transitively on the hyperboloid, confirming its maximal symmetry.

While the manifold is globally defined by the hyperboloid, physical calculations often require specific coordinate charts. There are several well-known coordinate patches that cover the whole or just a part of de Sitter space, such as global coordinates (covering the full manifold) or static coordinates (covering the region accessible to a single observer). For cosmological applications, we consider the Poincar\'e patch (or planar patch). This coordinate system covers only half of the de Sitter manifold—specifically, the expanding phase—and renders the metric in a conformally flat form.The embedding for the Poincaré patch is given by the following functions, with conformal time $\eta < 0$ and spatial coordinates $\mathbf{x} \in \mathbb{R}^{d-1}$:
\begin{equation}
\begin{aligned}
X^0(\eta,\mathbf{x}) &= \frac{H^{-2} - \eta^2 + |\mathbf{x}|^2}{-2\eta}\\ 
X^i (\eta, \mathbf{x}) &= -\frac{x^i}{H\eta}\\
%X^d(\eta,\mathbf{x}) & = \frac{H^{-2} + \eta^2 - |\mathbf{x}|^2}{-2\eta}
\end{aligned}
\end{equation}
where the spatial norm is defined by $|\mathbf{x}| \equiv \sqrt{\mathbf{x}\cdot\mathbf{x}} \equiv \sqrt{\delta_{ij} x^i x^j}$. It is crucial to note that this patch is geodesically incomplete; it only covers the region where $X^0 + X^d = 1/(-\eta H^2) > 0$. The boundary $\eta \to -\infty$ corresponds to the past null infinity of this patch, while $\eta \to 0$ represents the future spacelike infinity.

Substituting these embedding functions into the ambient Minkowski metric yields the induced metric:
\begin{equation}
ds^2 = -dt^2 + a^2(t) d\mathbf{x}\cdot d\mathbf{x} = a^2(\eta) (-d\eta^2 + d\mathbf{x}\cdot d\mathbf{x})
\end{equation}
where the scale factor in cosmic time $t$ is $a(t) = e^{Ht}$. The relationship between cosmic time $t$ and conformal time $\eta$ is given by:
\begin{equation}
\eta = -\frac{1}{H}e^{-Ht} \quad \implies \quad a(\eta) = -\frac{1}{H\eta}
\end{equation}
This metric form explicitly demonstrates that the Poincar\'e patch of de Sitter space is conformally equivalent to flat Minkowski space, a property that significantly simplifies the calculation of conformal field theories and propagators. The spacetime possesses a symmetry of time translation combined with a spatial scaling, characteristic of the dilation isometry $t \to t + t_0, \mathbf{x} \to e^{H t_0}\mathbf{x}$.

\bibliography{references}

\end{document}